\documentclass[a4paper]{jpconf}
\usepackage{graphicx,color}
\usepackage{amssymb}

\newcommand{\re}{RFeAsO}
\newcommand{\la}{LaFeAsO}
\newcommand{\ce}{CeFeAsO}
\newcommand{\pr}{PrFeAsO}
\newcommand{\sm}{SmFeAsO}
\newcommand{\gd}{GdFeAsO}
\newcommand{\laf}{LaFeAsO$_{1-x}$F$_x$}

\newcommand{\tn}{$T_{\rm{N}}$}
\newcommand{\ts}{$T_{\rm{S}}$}

\newcommand{\tnmu}{$T_{\rm{N}}^{\mu}$}
\newcommand{\tsrho}{$T_{\rm{S}}^{\rho}$}

\newcommand{\tcemu}{$T_{\rm{N}}^{Ce,\mu}$}
\newcommand{\tsmmu}{$T_{\rm{N}}^{Sm,\mu}$}
\newcommand{\tprmu}{$T_{\rm{N}}^{Pr,\mu}$}

\newcommand{\tnrmu}{$T_{\rm{N}}^{R,\mu}$}
\newcommand{\tnr}{$T_{\rm{N}}^{R}$}
\newcommand{\tnpr}{$T_{\rm{N}}^{Pr}$}
\newcommand{\tngd}{$T_{\rm{N}}^{Gd}$}

\newcommand{\figref}[1]{fig.\,\protect\ref{#1}}

\begin{document}
\title{Thermal expansion of RFeAsO (R=La,Ce,Pr,Sm,Gd)}

\author{R.~Klingeler, L.~Wang, U.~K{\"o}hler, G.~Behr, C.~Hess, B.~B\"uchner}
\address{Institute for Solid State Research, IFW Dresden, D-01171 Dresden, Germany}
\ead{r.klingeler@ifw-dresden.de}

\begin{abstract}
We present measurements of the thermal expansion coefficient $\alpha$ of polycrystalline RFeAsO (R
= La,Ce,Pr,Sm,Gd). Anomalies at the magnetic ordering transitions indicate a significant
magneto-elastic coupling and a negative pressure dependence of \tn . The structural transitions
are associated by large anomalies in $\alpha$. Rare earth magnetic ordering in \ce , \pr , and
\sm\ yields large positive anomalies at low temperatures.
\end{abstract}


Layered $\rm Fe_2As_2$-materials have raised enormous attention
due to the discovery of superconductivity with transition
temperatures $T_\mathrm{C}$ up to 28~K in \laf
.~\cite{Kamihara2008} Upon substitution of La by rare earths,
$T_\mathrm{C}$ is increased to above
50~K.~\cite{Chen2008a,Cheng2008,Ren2008c,Hess2009} Interestingly,
evolution of superconductivity is associated to the suppression of
a magnetically ordered orthorhombic phase, which has been found in
the undoped parent compound.~\cite{Luetkens2009} In \re , both
tetragonal distortion and magnetic ordering are observed at
intermediate temperatures around $\sim 150$\,K. A spin density
wave (SDW)-type of antiferromagnetic order evolves slightly below
the temperature \ts\ of orthorhombic distortion of the tetragonal
high temperature
phase.~\cite{Cruz2008,Klauss2008,Drew2008a,Maeter09}

Here, we present thermal expansion data of polycrystalline RFeAsO with R = La,Ce,Pr,Sm,Gd. Our
measurements yield a very sensitive measure of the volume changes of the materials. We find clear
anomalies of the coefficient of linear thermal expansion $\alpha$ at the structural and magnetic
transitions, i.e. at \ts\ and \tn , respectively. It has been shown earlier for \la , that
anomalous contributions to $\alpha$ are visible far above \ts .~\cite{Wang09} Similar effects are
found in \re\ with R = Ce,Pr,Sm,Gd. In addition, magnetic ordering of the rare earth moments is
accompanied by low temperature anomalies of the thermal expansion coefficient.


Preparation and characterization of the polycrystalline samples
has been described in Ref.~\cite{Kondrat2009}. The crystal
structure and the composition were confirmed by powder x-ray
diffraction. In addition, our samples have been characterized by
means of specific heat, magnetization, transport, and $\mu$SR
experiments.~\cite{Maeter09,Kondrat2009,Klingeler2008} For the
thermal expansion measurement a three-terminal capacitance
dilatometer was utilized, which allows an accurate study of sample
length changes. We measured the macroscopic length $L(T)$ of the
samples and calculated the coefficient of linear thermal expansion
$\alpha = 1/L \cdot dL/dT$, which is the first temperature
derivative of $L(T)$. For our polycrystalline samples the volume
expansion coefficient $\beta$ is given as $\beta = 3 \alpha$.


\begin{figure}
 \centering
 \includegraphics[width=0.9\columnwidth,clip]{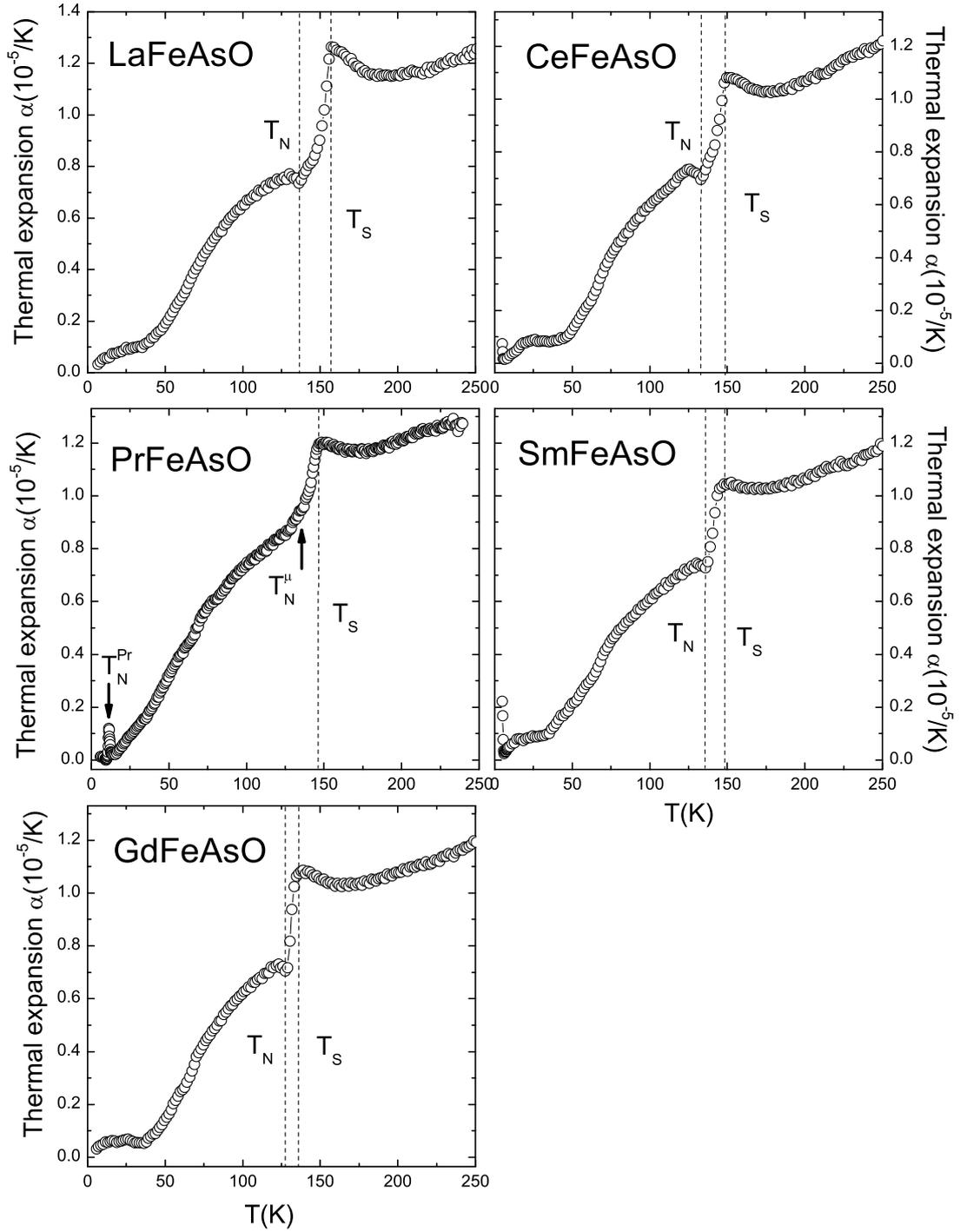}
 \caption{Temperature dependence of the coefficient of linear
thermal expansion, $\alpha (T)$, of RFeAsO (R = La,Ce,Pr,Sm,Gd). Two anomalies indicated by the
dashed lines are associated to a structural distortion at \ts\ and SDW-formation at \tn . For \pr
, no clear anomaly can be attributed to \tn . The arrow indicates \tprmu\ taken from $\mu$SR
data.~\cite{Maeter09} }
 \label{fig1}
\end{figure}

Figure~\ref{fig1} shows the linear thermal expansion coefficient $\alpha$ of \re\ with
R=La,Ce,Pr,Sm,Gd, between 5~K and 250~K. For all R (except Pr), the thermal expansion coefficient
exhibits two huge anomalies with opposite sign. The anomalies in $\alpha (T)$ can be attributed to
the structural and SDW transitions of the compound. The transition temperatures determined from
the positions of the extrema are marked by the dashed lines and are listed in Table~\ref{table}.

The SDW formation at $T_\mathrm{N}$ generates negative anomalies
in the thermal expansion coefficients. Note, that for \pr\ the
SDW-anomaly is not visible although the onset of magnetic order at
\tnmu = 137\,K has been demonstrated in $\mu$SR
studies.~\cite{Maeter09} According to the Ehrenfest relation, the
negative anomalies in $\alpha (T)$ at $T_\mathrm{N}$ qualitatively
imply a negative hydrostatic pressure dependence of \tn . This
finding is in agreement with resistivity studies on
LaFeAsO.~\cite{F4-08-7} The anomalies in $\alpha$ at
$T_\mathrm{N}$ indicate a strong coupling of the magnetic
transition to the crystal lattice. However, the shape of the
anomalies deviates from what is expected for second-order phase
transitions, probably due to the closeness to \ts .

\begin{table}
\caption{\label{table}Magnetic and structural transition temperatures of \re\ with R =
La,Ce,Pr,Sm,Gd as deduced from Figures \ref{fig1} and \ref{fig2}. For comparison, transition
temperatures from $\mu$SR (Ref.~\cite{Maeter09}) and resistivity studies (Ref.~\cite{Kondrat2009})
are listed, too.}
\begin{center}
\begin{tabular}{ccccccc}
\br
R&\tn &\ts & \tnr & \tnmu &\tsrho &\tnrmu \\
\mr
  La & (137$\pm$ 1)\,K & (157$\pm$ 1)\,K & -                & 139\,K & 158\,K & - \\
  Ce & (134$\pm$ 2)\,K & (148$\pm$ 2)\,K & $\lesssim 5$\,K  & 137\,K & 151\,K & 4.4\,K \\
  Pr & -               & (147$\pm$ 5)\,K & (11.3$\pm$ 0.3)\,K & 123\,K & 136\,K & 11\,K \\
  Sm & (136$\pm$ 2)\,K & (148$\pm$ 5)\,K & $\lesssim 5$\,K  & 138\,K & 160\,K & 4.7\,K \\
  Gd & (128$\pm$ 2)\,K & (136$\pm$ 5)\,K & - & -      & -      & - \\
\br
\end{tabular}
\end{center}
\end{table}

In contrast, the structural transition at $T_\mathrm{S}$ gives rise to a positive anomaly in
$\alpha$. Remarkably, this anomaly is very broad, extending to temperatures far above
$T_\mathrm{S}$. In particular, it has been shown previously for \la\ that the anomalous
contributions to $\alpha$ extend to significantly higher temperatures than the corresponding
anomalies found in specific heat, magnetization, and resistivity. The enhanced $\alpha$ suggests
the presence of strong fluctuations preceding the structural transitions at $T_\mathrm{S}$. So
far, the origin of these fluctuations is unknown. One might attribute them to a competing
instability in vicinity of the actual ground state. A possible scenario is a competing
orthomagnetic phase which was suggested in Ref.~\cite{Lorenzana2008}. In this scenario, long range
order of the competing magnetic phase is hindered by the orthorhombic distortion, whereas the
increase of the corresponding anomalous positive contribution to the thermal expansion coefficient
is truncated by the structural transition at \ts.

\begin{figure}[h]
\includegraphics[width=19pc]{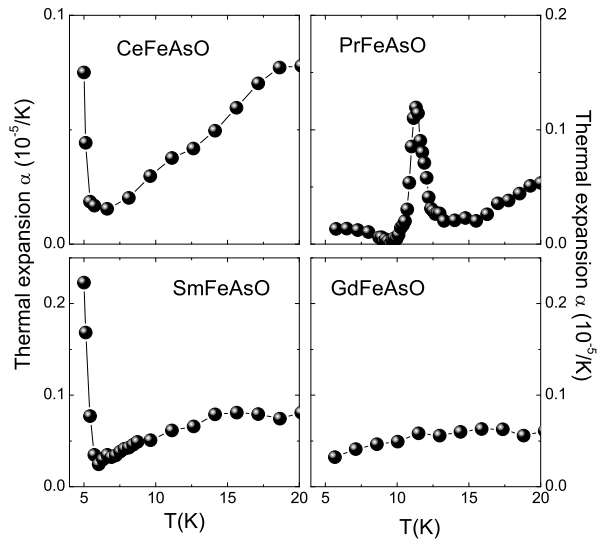}\hspace{2pc}%
\begin{minipage}[b]{17pc}\caption{\label{fig2}Temperature dependence of the coefficient of linear
thermal expansion, $\alpha (T)$, of RFeAsO (R = Ce,Pr,Sm,Gd) at temperatures below $T=20$\,K where
magnetic ordering of the rare earth moments is expected. Note, that the experimental setup
provides data for $T\geq 5$\,K, i.e. the complete anomaly is only visible for \pr\ which exhibits
\tnpr = 11.3\,K.}
\end{minipage}
\end{figure}

In the materials with magnetic R-sites, magnetic ordering of the
rare earth moments is found (see, e.g.,
\cite{Kimber08,Zhao2008natmat,Maeter09}). The evolution of rare
earth magnetic order is accompanied by strong volume changes.
Coupling of magnetic and lattice degrees of freedom is clearly
visible in \figref{fig2}. For \pr , there is a pronounced peak of
the thermal expansion coefficient at \tnpr = (11.3$\pm$0.3)\,K.
The observed ordering temperature agrees to previous neutron and
$\mu$SR data.~\cite{Kimber08,Maeter09}. Qualitatively, the strong
positive anomaly implies a positive hydrostatic pressure
dependence of \tnpr . Also for \ce\ and \sm , the data in
\figref{fig2} indicate a strong positive anomaly in $\alpha$
slightly below 5\,K, which is the lower temperature limit of our
device. In contrast, no anomaly is seen for \gd . Note, that
magnetic R-site ordering occurs in our samples at \tcemu =4.4\,K,
\tsmmu =4.7\,K, and \tngd = 3.7\,K.~\cite{Maeter09}.

In conclusion, our thermal expansion studies have been shown being
a sensitive probe for structural changes as well as Fe and rare
earth magnetic ordering in RFeAsO (R = La,Ce,Pr,Sm,Gd). The
magnetic and structural ordering phenomena are associated to large
anomalies in $\alpha$, which allow to determine the phase diagram.
Our data imply a negative pressure dependence of the Fe-ordering
transition and a positive one for Pr,Ce, and Sm ordering. Strong
fluctuations at $T\gg T_\mathrm{S}$ indicate a competing, possibly
magnetic instability to the ground state.

\ack We thank M. Deutschmann, S. M\"uller-Litvanyi, R. M\"uller, J. Werner, and S. Ga{\ss} for
technical support. Work was supported by the DFG through FOR 538 and project BE1749/12.

\section*{References}

\providecommand{\newblock}{}

\end{document}